\documentclass[12pt]{article}
\usepackage{color}
\usepackage{amssymb}
\usepackage{amsmath}
\usepackage{amsthm}
\usepackage[round]{natbib}
\usepackage{graphicx}
\usepackage{eurosym}
\usepackage{amsfonts}
\usepackage{multirow}
\usepackage{paralist}
\usepackage{verbatim}
\usepackage{subfig}
\usepackage{mathrsfs}
\usepackage{bm}
\usepackage{hyperref}
\usepackage{tabularx}
\usepackage{booktabs}
\usepackage{endnotes}
\usepackage[linesnumbered,lined,boxed,commentsnumbered]{algorithm2e}
\usepackage{soul}
\usepackage{enumitem}
\usepackage{mathtools}
\usepackage{geometry}
\usepackage{longtable}
\usepackage{graphicx}
\usepackage{subcaption}
\usepackage{float}
\usepackage{pgfplots}

\pgfplotsset{compat=1.15}  
\newtheorem{theorem}{Theorem} [section]

\newtheorem{corollary}[theorem]{Corollary}

\newtheorem{definition}[theorem]{Definition}
\newtheorem{example}[theorem]{Example}

\newtheorem{proposition}[theorem]{Proposition}

\allowdisplaybreaks[1]

\begin{document}

\title{The Shapley value and the strength of weak players in Big Boss games}
\author{Luis A. Guardiola\thanks{%
Departamento de Matemáticas, Universidad de Alicante,
03080, Spain. e-mail: luis.guardiola@ua.es} $^{,} $\thanks{%
Corresponding author.} \ and Ana Meca\thanks{%
I. U. Centro de Investigación Operativa, Universidad Miguel Hernández de Elche,
03202 Elche, Spain. e-mail: ana.meca@umh.es}}
\maketitle

\begin{abstract}
Big Boss Games represent a specific class of cooperative games where a single veto player, known as the Big Boss, plays a central role in determining resource allocation and maintaining coalition stability. In this paper, we introduce a novel allocation scheme for Big Boss games, based on two classical solution concepts: the Shapley value and the $\tau$-value. This scheme generates a coalitionally stable allocation that effectively accounts for the contributions of weaker players. Specifically, we consider a diagonal of the core that includes the Big Boss's maximum aspirations, the $\tau$-value, and those of the weaker players. From these allocations, we select the one that is closest to the Shapley value, referred to as the Projected Shapley Value allocation (PSV allocation). Through our analysis, we identify a new property of Big Boss games, particularly the relationship between the allocation discrepancies assigned by the $\tau$-value and the Shapley value, with a particular focus on the Big Boss and the other players. Additionally, we provide a new characterization of convexity within this context. Finally, we conduct a statistical analysis to assess the position of the PSV allocation within the core, especially in cases where computing the Shapley value is computationally challenging.

\bigskip \noindent \textbf{Key words:} Big Boss games, Shapley value, $\tau $-value, stable allocations

\noindent \textbf{2000 AMS Subject classification:} 91A12, 90B99
\end{abstract}

\newpage

\section{Introduction}

In the literature, games with a single dominant player have been widely studied in various economic contexts. A notable example is the market for information goods, where a single agent possesses the information while multiple demanders seek access to it \citep{muto1986information,nakayama1986bargaining,muto1989information}. Within this line of research, Big Boss Games \citep{muto1988big} represent a particular class of cooperative games in which there is a single veto player, known as the big boss, whose presence decisively influences resource allocation and coalition stability. The remaining players, referred to as weak players, are subject to the strategic dominance of the big boss.

Several studies have analyzed the properties of Big Boss games in different contexts. In particular, \cite{tijs1990big} examines Big Boss games, clan games, and information market games. This work defines these game classes, analyzes their economic foundations, and highlights key examples, such as markets with a dominant seller, landlord-worker models, and bankruptcy problems. It also identifies information market games as a special case of Big Boss games, where the Big Boss is the initially informed trader.  Expanding on this framework, \cite{bahel2016core} generalizes clan games and Big Boss games under the concept of generalized Big Boss games or veto games, characterizing their core and establishing its equivalence with the bargaining set. Additionally, \cite{liu2021dynamic} investigates the effect of veto power on players' income in a multi-player dynamic bargaining game.  

Big Boss games have also been studied from the perspective of interval games in \cite{alparslan2011big}. An extension of this work is presented in \cite{gok2023big}, where each coalition value is modeled as a fuzzy interval of real numbers rather than a simple interval. Moreover, more recently, \cite{ozcan2024big} has also explored Big Boss games within the framework of situations involving partial information, also known as cooperative grey games.

A key characteristic of Big Boss Games is that the core forms a parallelotope, in the sense that the core imputation of each weak player is bounded between zero and their marginal contribution to the grand coalition. Moreover, the $\tau $-value coincides with the nucleolus, and both are located at the center of the core. Additionally, the Shapley value of the big boss is generally lower than the $\tau $-value and the nucleolus, with equality occurring if and only if the game is convex. The $\tau $-value appears to play the role of a solomonic allocation, balancing the maximal aspirations of the big boss and the weak players on the core.

In this paper, we propose an alternative allocation for Big Boss
games. Our approach builds on two well-known cooperative game theory
allocation rules: the Shapley value and the tau-value. While the Shapley value
ensures a fair division of the overall gains based on each player's marginal
contribution, and the tau-value focuses on coalitional stability, both
allocation rules can overlook the nuanced contributions of weaker players in such
hierarchical game structures.

To address this, we introduce an allocation method that is not only
coalitionally stable but also explicitly accounts for the role of weak
players, ensuring that their contributions to the game are reflected in the
final allocation. Specifically, we focus on a subset of the core
referred to as the $\tau $-diagonal that includes $\tau $-value, the Big Boss's
maximal aspirations while ensuring that weak players receive a consistent
percentage of their marginal contribution to the grand coalition. Within
this subset, we identify the allocation that is closest to the Shapley
value, which we call the Projected Shapley Value allocation (henceforth, PSV
allocation). Through this study, we introduce a novel property that establishes a relationship between the relative positions of the $\tau $-value and the Shapley value within the core.

Furthermore, we investigate the statistical properties of the PSV
allocation, particularly in cases where calculating the exact Shapley value
is computationally challenging. This aspect is crucial in large-scale games
where the complexity of traditional methods can be prohibitive.

The outline of the paper is as follows. We start by introducing
some preliminary concepts in Section \ref{intro}. In Section \ref{big}, we define the concept of the $\tau $-diagonal and the PSV allocation. Additionally, we establish a novel property for this class of games, demonstrating that the difference between the $\tau $-value and the Shapley value for the Big Boss can never be smaller than the corresponding difference for any weak player. This property allows us to determine a feasible range within the $\tau $-diagonal for the projection of the Shapley value. Furthermore, we provide a new characterization of convexity based on the position of the PSV allocation. In Section \ref{esta}, we present a statistical study that facilitates the identification of the PSV allocation's position as a function of the game’s size. This approach ensures that for large-scale games, the Shapley value does not need to be explicitly computed.
Section \ref{con} concludes. 

\section{Preliminaries cooperative game theory} \label{intro}

A cooperative game with transferable utility (TU-game) consists of a set of
players $N=\{1,...,n\}$ and the characteristic function $v\text{, which
corresponds}$ to each subset of the set $N$ with a number from the set of
real numbers. The subsets formed by the set $N$ are called coalitions, which
are denoted by $S$. Formally the characteristic function is an application $%
v :2^{N} \longrightarrow \mathbb{R}$ such that $v(\emptyset )=0$. The value $%
v(S)$ of the characteristic function measures the maximum benefit that the
members of the coalition $S$ can achieve by cooperating together. The
coalition formed by all agents, $N$, it is called the grand coalition. We
denote by $s$ the cardinality of the set $S\subseteq N$, i.e. $\left\vert S\right\vert=s$.


One of the key issues addressed by cooperative game theory is how to allocate the total profit generated by the grand coalition once it has been formed. This allocation is represented by a vector \( x \in \mathbb{R}^{n} \), where \( n \) denotes the number of players in the set \( N \). The class of superadditive games is particularly notable, as it encourages the formation of the grand coalition to maximize the total profit. Formally, a TU-game \( (N,v) \) is superadditive if for every two coalitions \( S, T \subseteq N \) with \( S \cap T = \emptyset \), it holds that \( v(S \cup T) \geq v(S) + v(T) \). 



Cooperative game theory provides several approaches for allocating the gains resulting from collaboration. These solutions can be classified into two categories: set solutions and point solutions. Set solutions involve identifying the allocations that meet specific conditions by excluding those that do not. In contrast, point solutions are determined through an axiomatic characterization, meaning they represent the unique allocation that satisfies a given set of properties.

The core of a TU-game is the most significant set solution. It includes all efficient allocations that are coalitionally stable, meaning no coalition has an incentive to break away from the grand coalition without reducing its own payoff. Formally,

\begin{equation*}
Core(N,v)=\left\{ x\in \mathbb{R} ^{n}:\sum_{i\in
N}x_{i}=v(N)\text{ and, for all }S\subset N, \sum_{i\in S}x_{i}\geq v(S)
\right\} .
\end{equation*}


The result presented by \cite{bondareva1963some} and \cite{shapley1967balanced} provides a necessary and sufficient condition for the non-emptiness of the core of a TU-game. Specifically, one of the key theorems in cooperative game theory asserts that a TU-game has a non-empty core if and only if it is balanced.


A point solution $\varphi$ refers to a function that, for each TU-game $%
(N,v) $, determines an allocation of $v(N)$. Formally, we have $%
\varphi:G^{N}\longrightarrow \mathbb{R}^{n}$, where $%
G^{N}$ denotes the class of all TU-games with player set $N$, and $%
\varphi_{i}(v)$ represents the profit assigned to player $i\in N$ in the
game $v\in G^{N}$. Therefore, $\varphi(v)=(\varphi_{i}(v))_{i \in N}$ is a
profit vector or allocation of $v(N)$. For a comprehensive overview of
cooperative game theory, we recommend referring to \cite%
{gonzalez2010introductory}.

The Shapley value, first introduced in \cite{shapley1953value}, is a widely recognized single-valued solution in cooperative game theory. The Shapley value of convex games always belongs to the core and is the barycenter of the core (see \cite{shapley1971cores}). Moreover, it is a linear operator on the set of all TU games. For a profit game \( (N,v) \), \( \phi \) is defined as \( \phi(N,v) = (\phi_i(N,v))_{i \in N} \), where for each \( i \in N \),

\begin{equation*}
\phi _{i}(v)=\sum\limits_{S\subseteq N\backslash \{i\}}\frac{s!(n-s-1)!}{n!}%
\cdot \left[ v(S\cup \{i\})-v(S)\right] .
\end{equation*}

The nucleolus \citep{schmeidler1969nucleolus}, denoted as \( \eta(v) \), is a solution concept in cooperative game theory that seeks to identify the most stable allocation within the core. It is determined by addressing the maximum dissatisfaction of all coalitions, where dissatisfaction is quantified as the excess, which is the difference between the value of a coalition and the sum of the payoffs distributed to its members. The nucleolus minimizes the largest excess, ensuring that the allocation is as fair as possible from the perspective of each coalition's dissatisfaction.

Given a cooperative game $(N,v)$ with a nonempty core, the \( \tau \)-value \citep{tijs1981bounds} \( \tau(v) = (\tau_i(v))_{i \in N} \) is defined as the convex combination of the marginal contributions and the maximum residual contributions of the players. Specifically, for each \( i \in N \), the marginal contribution is given by \( M_i(v) = v(N) - v(N \setminus \{i\}) \), and the maximum residual contribution is defined as \( m_i(v) = \max \{ v(S) - \sum_{j \in S} M_j(v) : S \subseteq N, i \notin S \} \). Letting \( M(v) = (M_i(v))_{i \in N} \) and \( m(v) = (m_i(v))_{i \in N} \), the \( \tau \)-value is the unique allocation satisfying \( \sum_{i \in N} \tau_i(v) = v(N) \).

A cooperative game $(N,v)$ is a big boss game if there is one player, denoted by $b^{v}$, statisfying the followings conditions:

\begin{itemize}
\item[(B1)] The game is monotonic, i.e., $v(S)\leq v(T)$ for all $S,T\subseteq
N$ with $S\subseteq T.$

\item[(B2)] There is a player $b^{v}\in N$ such that $v(S)=0$ if $b^{v}\notin
N $.

\item[(B3)] If $b^{v}\in S,$ then $v(N)-v(S)\geq \sum_{i\in N\setminus
S}M_{i}(v).$
\end{itemize}

(B2) establishes that there exists a dominant player, denoted as $b^{v}$, such that any coalition excluding $b^{v}$ has no worth. Meanwhile, (B3) ensures that the total contribution of any coalition lacking $b^{v}$ to the grand coalition is at least as large as the sum of the individual contributions of its members. Consequently, weak players can enhance their influence by collaborating and forming coalitions. It is important to note that a Big Boss game $v$ satisfies superadditivity, which follows directly from the monotonicity (B1) and condition (B2). We denote a class of all Big Boss games with $n$ players as $BBG^{N}.$ 

The core of a Big Boss game is always nonempty and takes the form of a simple parallelepiped structure.

\begin{equation*}
Core(N,v):=\left\{ x\in \mathbb{R}^{n }:\sum_{i\in
N}x_{i}=v(N),0\leq x_{i}\leq M_{i}(v)\text{ for all }i\in N\setminus
\{b^{v}\}\right\} .
\end{equation*}

Moreover, \cite{muto1988big} demonstrated that the \( \tau \)-value is the barycenter of the core of Big Boss games and can be explicitly calculated as follows:

\begin{equation*}
\tau (v):=\left\{ 
\begin{array}{ll}
\frac{1}{2}M_{i}(v), & i\in N\setminus \{b^{v}\}, \\ 
&  \\ 
v(N)-\frac{1}{2}\sum_{i\in N\setminus \{b^{v}\}}M_{i}(v), & i=b^{v}.%
\end{array}%
\right.
\end{equation*}

Furthermore, an interesting result shows the relationship between the nucleolus, the Shapley value, the \( \tau \)-value, and convexity:

\begin{theorem}[\cite{muto1988big}]\label{muto}
Let \( v \in BBG^{N} \). Then the following assertions are equivalent:

\begin{itemize}
\item[1.] $\phi _{ b^{v}}(v)=\tau_{ b^{v}}(v)=\eta_{ b^{v}}(v).$

\item[2.] $\phi (v)=\tau(v)=\eta(v).$

\item[3.] $v$ is convex.
\end{itemize}

\end{theorem}

\section{The proyected Shapley value for Big Boss games} \label{big}


Big Boss games can be understood as a conflict between the interests of the Big Boss and the weak players. If we focus our search for allocations within a stable framework such as the core, we can define the following elements of the core:
\medskip

$e_{0}(v):=\left\{ 
\begin{array}{ll}
0, & i\in N\setminus \{b^{v}\}, \\ 
&  \\ 
v(N), & i=b^{v},%
\end{array}%
\right. $ and $e_{1}(v):=\left\{ 
\begin{array}{ll}
M_{i}(v), & i\in N\setminus \{b^{v}\}, \\ 
&  \\ 
v(N)-\sum_{i\in N\setminus \{b^{v}\}}M_{i}(v), & i=b^{v},%
\end{array}%
\right. $
\medskip

The reader may notice that the allocations \( e_{0}(v) \) and \( e_{1}(v) \) are antagonistic. In \( e_{0}(v) \), the Big Boss receives the maximum possible benefit within the core, while in \( e_{1}(v) \), the Big Boss is allocated the minimum benefit possible. Additionally, \( e_{1}(v) \) can be interpreted as the maximum percentage of their marginal contribution that the weak players can receive, provided that this percentage is the same for all of them. Due to the characteristic structure of the core in Big Boss games, it is apparent that these allocations represent two extreme points of the core, and we can define a diagonal that connects them that we called $\tau$-diagonal.

\begin{definition}[$\tau$-diagonal]
Let $v\in BBG^{N}$. Define $T(N,v):=\left\{ \tau ^{\rho }(v):\rho \in
\lbrack 0,1]\right\} $ where,%
\begin{equation*}
\tau ^{\rho }(v):=\left\{ 
\begin{array}{ll}
\rho M_{i}(v), & i\in N\setminus \{b^{v}\}, \\ 
&  \\ 
v(N)-\rho \sum_{i\in N\setminus \{b^{v}\}}M_{i}(v), & i=b^{v},%
\end{array}%
\right.
\end{equation*}
\end{definition}

It is easy to see that this set is within the core since it can be expressed
as:%
\begin{equation*}
T(N,v):=\left\{ (1-\rho )\cdot e_{0}(v)+\rho \cdot e_{1}(v):\rho \in \lbrack
0,1]\right\} ,
\end{equation*}%

Note that $\tau ^{\frac{1}{2}}(v)=\frac{1}{2}\cdot e_{0}(v)+\frac{1}{2}\cdot
e_{1}(v)=\tau (v)$ and $T(N,v)\subseteq Core(N,v).$ It is important to note that there is a particular case: when all the marginal contributions of the weak players to the grand coalition are zero. In this case, the core is reduced to a single point ($e_{0}(v) = e_{1}(v) = \tau (v)$), which is of no further interest in this study.

Our next goal will be to find an allocation within $T(N,v)$ that fairly
compensates the weak players for their contribution to the grand coalition.
Since the $\tau$-value provides a Solomonic solution to this problem (50\%
of the marginal contributions to the weak players) and is also the geometric
center of the tau-diagonal, we ask ourselves whether other criteria exist to
improve the allocation that the $\tau$-value grants to the weak players. For
this reason, we will take the Shapley value $\phi (v)$ as a reference,
knowing that it is not always an allocation within the core for the class of
Big Boss games.

The following result shows us the existence and exact location within the
set $T(N,v)$ of the allocation that is closest to the Shapley value.

\begin{theorem}\label{dis}
Let $v\in BBG^{N}$. Then, the allocation from $T(N,v)$ that is at the
minimum distance from $\phi (N,v)$ is $\tau ^{\alpha ^{v}}(v)$ with $\alpha ^{v}:=\min \left\{ \rho^{v},1\right\}$ where:
\begin{equation*}
\rho^{v} := 
\begin{cases} 
\frac{\left( v(N) - \phi_{b^{v}}(v) \right) \cdot \left( \sum_{i \in N \setminus \{b^{v}\}} M_{i}(v) \right) + \sum_{i \in N \setminus \{b^{v}\}} \left( \phi_{i}(v) \cdot M_{i}(v) \right)}{\left( \sum_{i \in N \setminus \{b^{v}\}} M_{i}(v) \right)^2 + \sum_{i \in N \setminus \{b^{v}\}} M_{i}^2(v)}, & \text{if } \sum_{i \in N \setminus \{b^{v}\}} M_{i}>0 \\
1, & \text{otherwise. }
\end{cases}
\end{equation*}

\end{theorem}

\begin{figure}[htbp]
\begin{center}
\begin{tikzpicture}

    \draw[gray, thick] (0,3) -- (1.5,0) -- (0,-3) -- (-1.5,0) -- cycle;
    
    \node at (-2.5,-1) {{\color{gray}\textbf{Core($N$, $v$)}}};

    \draw[blue, very thick] (0,-3) -- (0,3);
    \filldraw[black] (0,3) circle (2pt); 
    \filldraw[black] (0,-3) circle (2pt); 
    
    \node[anchor=east] at (0,3) {$e_1(v)$};
    \node[anchor=east] at (0,-3) {$e_0(v)$};
    
    \filldraw[red] (0,0) circle (2pt); 
    \node[anchor=west] at (0,0) {$\tau(v)$};
    
    \filldraw[red] (3,2) circle (2pt); 
    \node[anchor=west] at (3,2) {$\phi(v)$};
    
    \draw[red, dashed] (3,2) -- (0,2);
    
    \filldraw[red] (0,2) circle (2pt); 
    \node[anchor=west] at (0,2) {$\tau^{\alpha^v}(v)$};

\end{tikzpicture}
\caption{Projection of the Shapley value onto the $\tau$-diagonal}
\label{fig1}
\end{center}
\end{figure}
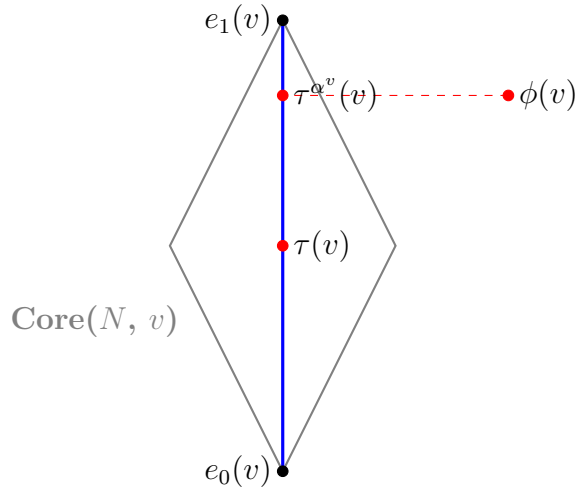

The Appendix provides the proofs of all results. We will now refer to $\tau ^{\alpha^{v} }(v)$ as the Projected Shapley Value
allocation (henceforth PSV allocation). Note that, $\alpha ^{v}$ can be
interpreted as the percentage that the PSV allocation assigns to the weak
players based on their marginal contributions. Figure \ref{fig1} shows a fictitious projection of the Shapley value onto the $\tau$-diagonal for a 3-player game.

The following example illustrates the calculation of the Shapley projection for a 3-player game.

\begin{example}
Consider a Big Boss game $(N,v)$ with $N = \{1, 2, 3\}$ and the
following coalition values: 
\begin{equation*}
\begin{array}{c|c|c|c|c|c|c|c|}
S & \{1\} & \{2\} & \{3\} & \{1,2\} & \{1,3\} & \{2,3\} & \{1,2,3\} \\ \hline
v(S) & 56 & 0 & 0 & 111 & 136 & 0 & 140
\end{array}
\end{equation*}

In this case, we have $b^{v}=1$,  $M(v) = (140, 4, 29)$, $e_{0}(v) = (140, 0, 0)$ and $e_{1}(v) = (107, 4, 29)$. It is straightforward to compute $\rho ^{v}=\alpha ^{v}=0.9324$, which leads to   $\tau ^{\alpha ^{v}}(v) = (109.23, 3.7297, 27.0403)$. Furthermore, the Shapley value and the $\tau$-value are given by  
\[
\phi(v) = (106.5, 10.5, 23), \quad \tau(v) = (123.5, 2, 14.5).
\]  
Considering the difference  
\[
\tau(v) - \phi(v) = (17, -8.5, -8.5),
\]  
we observe that the increase granted by the $\tau$-value to the Big Boss, compared to the Shapley value, is greater than the increase assigned to the weak players.

\end{example}

The previous example shows that the projection of the Shapley value is close to \( e_{1}(v) \).  We wonder if the PSV allocation can take any value within the segment \( [e_{0}(v), \tau(v)) \), that is, if it is possible that $\alpha^{v} < \frac{1}{2}$. The following example, based on a sample of Big Boss games, suggests that, for the case of three players, this is not possible.

\begin{example}
The set of players is $N = \{1, 2, 3\}$ and $v\in BBG^{N}$. We randomly generate 500 Big Boss games. We use uniform variables to assign a random natural value between \([1,100]\) to \( v(\{1\}) \). Subsequently, the values \( v(\{1,2\}) \) and \( v(\{1,3\}) \) take a value within the range \([v(\{1\}), v(\{1\})+100]\), and finally, \( v(N) \) take a value between \(\left[\max_{i \in N} v(\{1, i\}), 100+\max_{i \in N} v(\{1, i\}) \right]\). Moreover, we verify that all the games satisfy condition (B3) of the Big Boss.  
Figure \ref{fi1} shows a histogram with the values of \( \alpha^{v} \) obtained for each generated game.

\begin{figure}[H]
    \centering
    \includegraphics[width=0.5\linewidth]{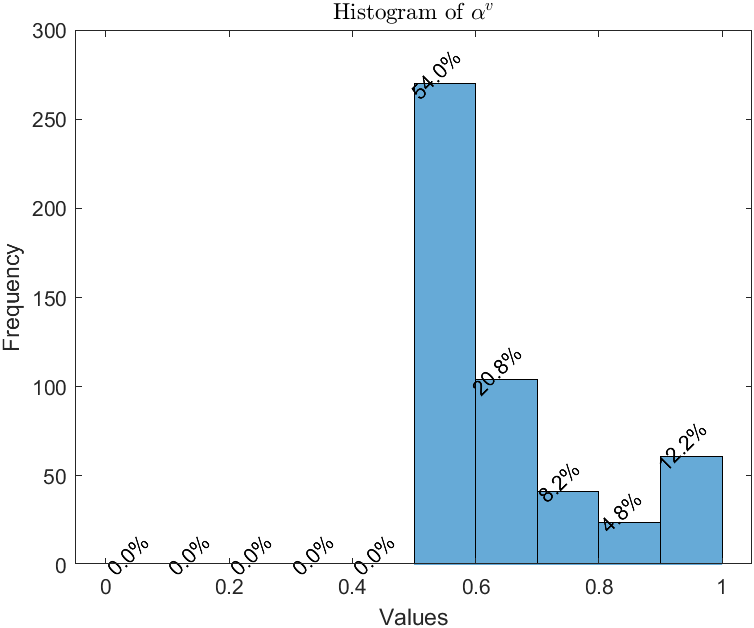}
\caption{Histogram of the \( \alpha ^{v} \) values for 3-player games}  
\label{fi1}
\end{figure}

\end{example}

The example indicates that the values of\( \alpha^{v} \) are consistently greater than or equal to $\frac{1}{2}$.  Our next goal is to prove that the PSV allocation always lies in the segment \( [\tau (v),e_{1}(v)] \). The following result provides a sufficient condition for this to hold.

\begin{proposition}\label{suf}
Let $v\in BBG^{N}$. If  $\tau _{b^{v}}(v)-\phi _{b^{v}}(v) \geq \max_{i \in N\setminus \{b^{v}\}} \{ \tau _{i}(v)-\phi _{i}(v)\}$, then  \( \alpha ^{v}\geq \frac{1}{2} \).
\end{proposition}

This property states that the increase granted by the $\tau$-value to the Big Boss, relative to the Shapley value, must be at least as large as the increase granted to any weak player. In other words, the tau-value should never compensate a weak player more than it compensates the Big Boss.

The following proposition provides an equivalent version of property (B3), which will be useful to us later.
\begin{proposition}\label{pa3}
Let \( v \in BBG^{N} \). Then, for any \( S \subseteq N \) such that \( b^{v} \in S \), the following holds:

$$\left( n-s -1\right) v(N)+v(S) \leq \sum_{i\in
N\setminus S}v(N\setminus \{i\}).$$

\end{proposition}

The main result of this paper demonstrates that the sufficient condition stated in Proposition \ref{suf} always holds in Big Boss games.

\begin{theorem}\label{te}
Let $v\in BBG^{N}$. Then, $\tau _{b^{v}}(v)-\phi _{b^{v}}(v) \geq \max_{i \in N\setminus \{b^{v}\}} \{ \tau _{i}(v)-\phi _{i}(v)\}$.
\end{theorem}

As the reader may observe, this property relies on two well-established allocation rules in the literature and introduces a novel property governing the relationship between the Shapley value and the \( \tau \)-value for Big Boss games. As an immediate consequence of this property, we have that the projection of the Shapley value always lies within the segment \( [\tau (v), e_{1}(v)] \), as demonstrated by the following corollary.

\begin{corollary}\label{col}
Let $v\in BBG^{N}$. Then, $\alpha ^{v}\geq \frac{1}{2}$
\end{corollary}

Theorem \ref{muto} characterizes the convexity of Big Boss games through the coincidence of the \( \tau \)-value and the Shapley value. We now ask whether convexity can also be characterized by the position that the PSV allocation occupies on the \( \tau \)-diagonal. The following result provides a necessary and sufficient condition for the convexity of a Big Boss game in terms of \( \alpha^{v} \) and the marginal contributions to the grand coalition of the weak players.

\begin{proposition}\label{conv}
Let $v\in BBG^{N}$ with \( M_i(v) > 0 \) for all \( i \in N\setminus \{b^{v}\} \). Then, $v$ is convex if and only if $\alpha ^{v}=\frac{1}{2}$
\end{proposition}

As we have just proven, the PSV allocation always provides the weak players with a percentage that is greater than or equal to the one assigned to them by the $\tau$-value. Moreover, it is the value closest to the Shapley value of \( T(N,v) \) and is coalitionally stable. In the next section, we will conduct a statistical study on the location of the PSV allocation within the $\tau$-diagonal.

\bigskip

\clearpage
\newpage

\section{Statistical analysis of the impact of increasing weak players} \label{esta}

In this section, we aim to analyze whether the parameter $\alpha^{v}$ exhibits a predictable behavior as the number of players increases. It is important to note that computing $\alpha^{v}$ requires knowledge of the Shapley value, whose calculation becomes computationally demanding for games with more than 15 players. For this reason, we investigate whether the values of this parameter, obtained from a family of Big Boss games with the same number of players, can be predicted—that is, whether they follow a specific probability distribution.

To achieve this goal, we first consider different numbers of players, specifically \( n = \{4, 5, 6, 7\} \), and define the population as the set of all possible Big Boss games. We extract a sample of 5000 such games, constructed as follows: all games take natural values. For each sample element, a seed value (\(\mu\)) is selected from a uniform distribution over the interval \([1, 100]\). For simplicity, we designate player 1 as the Big Boss. Next, the value of \( v(\{1\}) \) is randomly drawn from a uniform distribution over the interval \([1, \mu]\). For two-player coalitions that include the Big Boss, the value is selected from the interval \([v(\{1\}), v(\{1\}) + \mu]\). For three-player coalitions, the value is drawn from the interval \(\left[\max_{i \in N} v(\{1, i\}), \max_{i \in N} v(\{1, i\}) + \mu\right]\), and so on.

Once each game is generated, we verify that it satisfies all the necessary conditions to be classified as a Big Boss game. If it does, we compute the Shapley value and, finally, the value of the expression \( \rho^{v} \) (see (\ref{foru}) for more details). The reason for calculating  \(\rho^{v}\) instead of directly computing \(\alpha^{v}\) is crucial: when \(\alpha^{v}\) assigns a value of 1 to all instances of \(\rho^{v}\) that exceed 1, we lose relevant information about the exact point on the line containing the segment \([e_{0}(v), e_{1}(v)]\) where the Shapley value is projected.  

Thanks to Corollary \ref{col}, we know that the projection always occurs from the midpoint of this segment in the direction of \(e_{1}(v)\). However, in this case, we are particularly interested in measuring how many of these projections exceed the segment’s endpoint and by how much, in order to identify a probability density function that best fits the sample.

If we consider the populations of Big Boss games with 4 and 5 players, we obtain the following distribution of the values of \( \rho^{v} \):

\begin{figure}[H]
    \centering
    \begin{minipage}{0.45\textwidth}
        \centering
        \includegraphics[width=\linewidth]{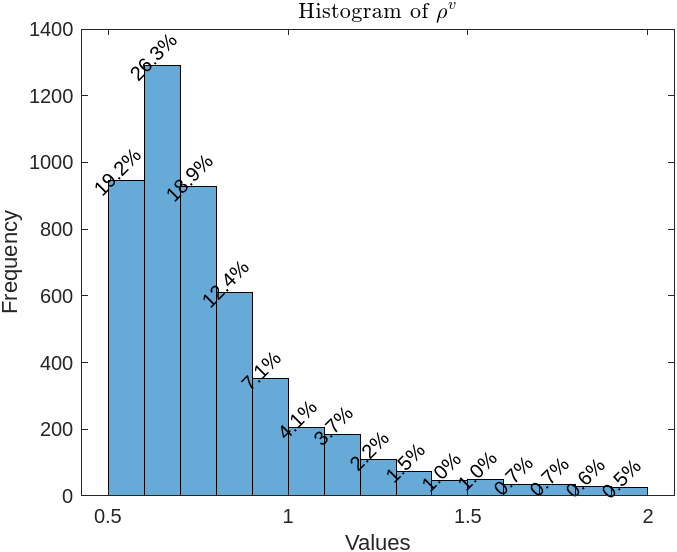}  
        \caption{Histogram for 4-player game}  
    \end{minipage}
    \hfill  
    \begin{minipage}{0.45\textwidth}
        \centering
        \includegraphics[width=\linewidth]{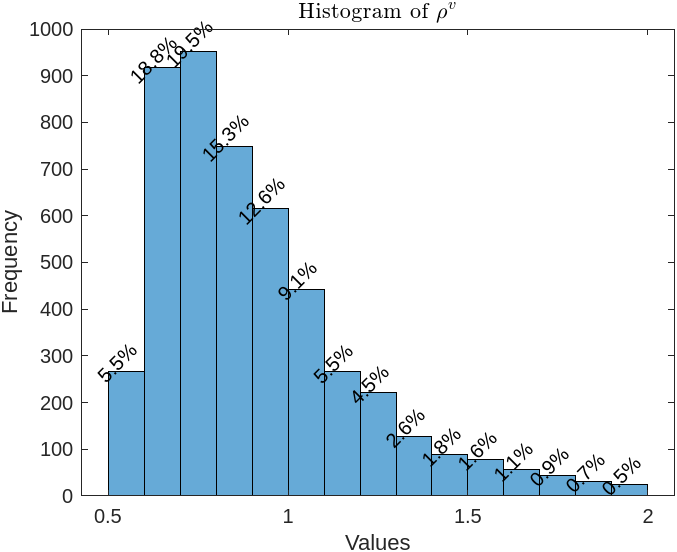}  
        \caption{Histogram for 5-player game}  
    \end{minipage}

\end{figure}

As we can see in the (unnormalized) histograms, the values of \( \rho^{v} \) appear to be concentrated around 0.5 in the 4-player game. However, in the 5-player case, they exhibit a rightward shift. For Big Boss games with 6 and 7 players, we obtain:

\begin{figure}[H]
    \centering
    \begin{minipage}{0.45\textwidth}
        \centering
        \includegraphics[width=\linewidth]{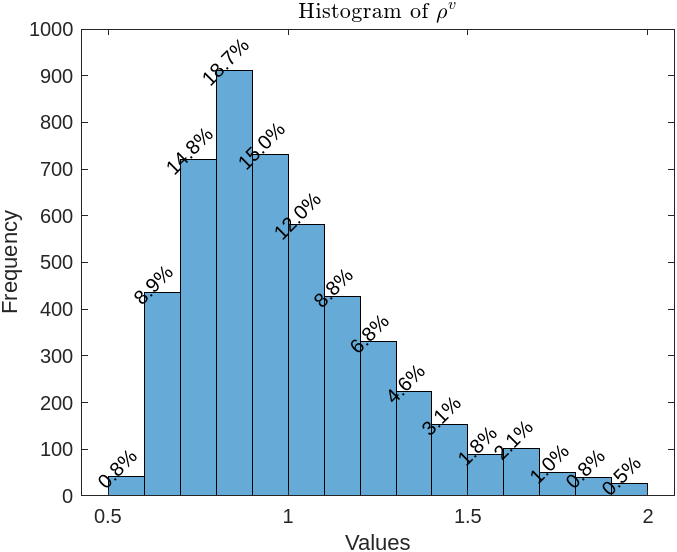}  
        \caption{Histogram for 6-player game}  
    \end{minipage}
    \hfill  
    \begin{minipage}{0.45\textwidth}
        \centering
        \includegraphics[width=\linewidth]{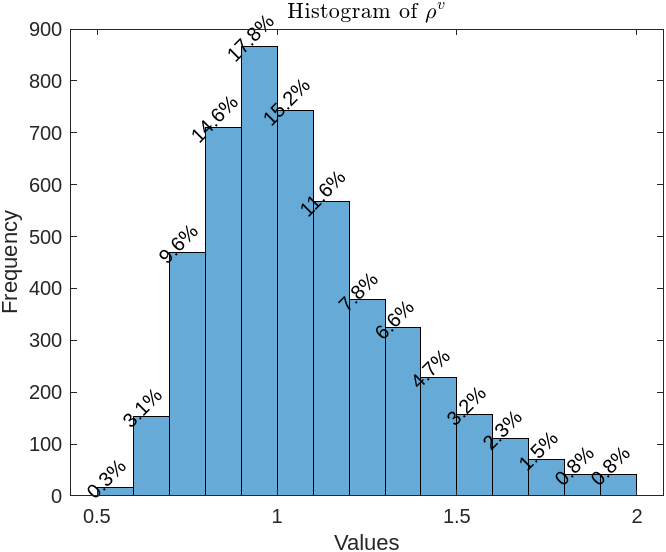}  
        \caption{Histogram for 7-player game}  
    \end{minipage}

\end{figure}

The histograms appear to resemble the shape of a Log-Normal density function. To verify this, we normalize the histogram and overlay a fitted log-normal density function, considering three samples from populations of Big Boss games with 4, 6, and 8 players.  

For the 4- and 6-player games, we take samples of size 5000, while for the 8-player game, the sample size will be 10000. This choice is due to the significantly larger number of coalitions in an 8-player game compared to those with 4 and 6 players. Additionally, we compare the cumulative distribution functions of the log-normal fit with the empirical distribution function of the obtained sample.

\begin{figure}[H]
    \centering
    
    \begin{minipage}{0.45\textwidth}
        \centering
        \includegraphics[width=\linewidth]{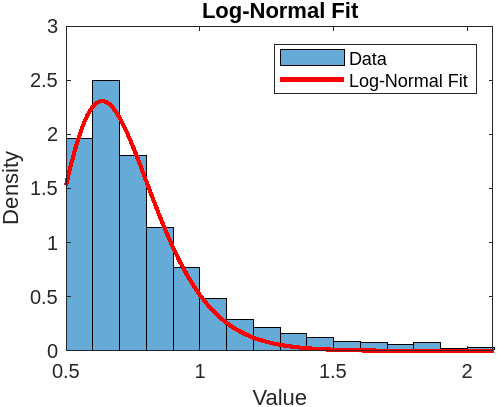}
    \end{minipage}
    \hfill
    \begin{minipage}{0.45\textwidth}
        \centering
        \includegraphics[width=\linewidth]{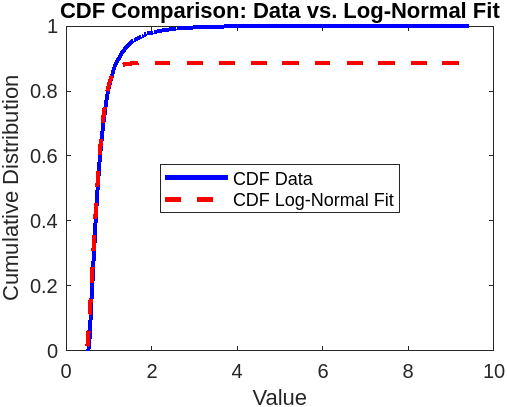}
    \end{minipage}

    \vspace{0.5cm}  

    \begin{minipage}{0.45\textwidth}
        \centering
        \includegraphics[width=\linewidth]{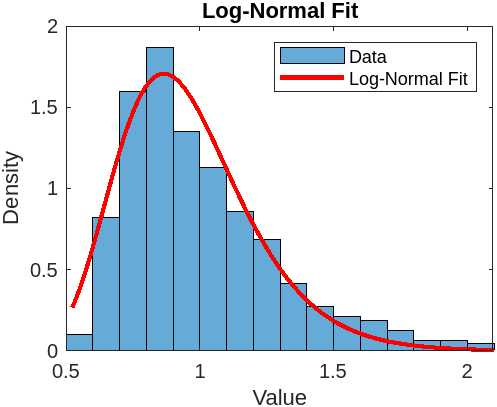}
    \end{minipage}
    \hfill
    \begin{minipage}{0.45\textwidth}
        \centering
        \includegraphics[width=\linewidth]{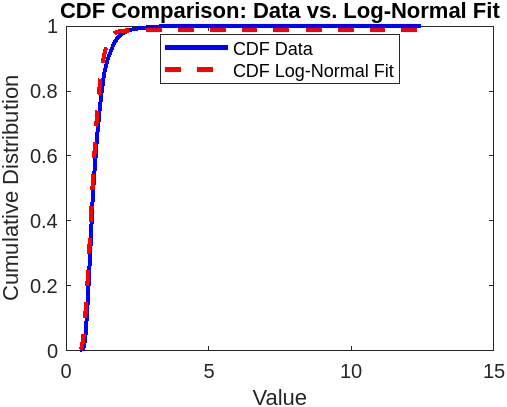}
    \end{minipage}

    \vspace{0.5cm}  

    \begin{minipage}{0.45\textwidth}
        \centering
        \includegraphics[width=\linewidth]{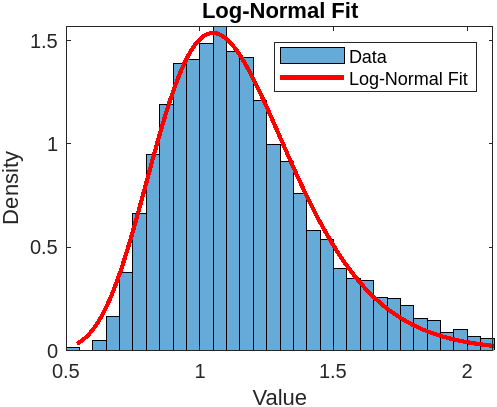}
    \end{minipage}
    \hfill
    \begin{minipage}{0.45\textwidth}
        \centering
        \includegraphics[width=\linewidth]{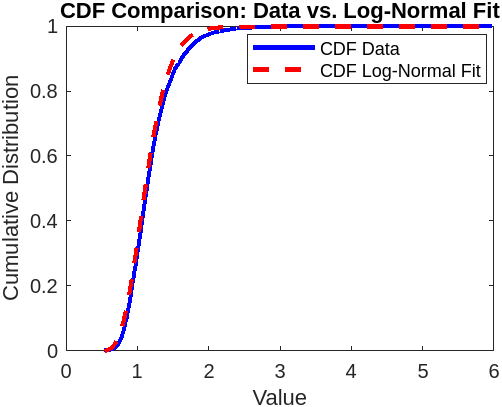}
    \end{minipage}

    \caption{Comparison of the density function and distribution of Log-Normal and samples for games of size 4, 6, and 8.}
    \label{fig:comparison}
\end{figure}

It is important to highlight that as the number of players increases, both the normalized histogram and the distribution functions of the samples and the Log-Normal random variable become more similar. To gain greater confidence in the similarity of the data, we performe a Kolmogorov-Smirnov test to verify that the values \( \rho^{v} \) follow a Log-Normal distribution with the parameters obtained through least squares fitting. This test is a hypothesis test where the null hypothesis states that the data follow the Log-Normal distribution, and the alternative hypothesis is the opposite. To accept the null hypothesis, a p-value greater than 0.05 is required. Table \ref{tab:esta} presents the results for the p-value and the Kolmogorov-Smirnov statistic (\( KS \)), which should be close to zero (a value of 0.05 or less indicates a good fit). It is well-known that if the sample size is very large, the goodness-of-fit test may reject the null hypothesis; therefore, we consider a sample of 500 Big Boss games with different numbers of players.

\begin{table}[h!]
\centering
\begin{tabular}{|c|c|c|}
$n$ & $p$-value & $KS$   \\
\hline
 4 & 0.000006  &  0.1122  \\
\hline
 6 &  0.1992  &  0.0477 \\
\hline
 8 &  0.0869  &  0.0557   \\
\hline
 10 &  0.5964  &  0.0340   \\
\hline
\end{tabular}
\caption{Kolmogorov-Smirnov test results}
\label{tab:esta}
\end{table}

Therefore, from this point on, we can consider that the population of parameters \( \rho^{v} \) obtained from Big Boss games follows this probability distribution. Formally, Let \(X\) be a random variable that follows a Log-Normal distribution, denoted as:
\[
X \sim \text{Log-Normal}(\mu, \sigma^2)
\]
which means that its probability density function is given by:

\[
f_X(x) = \frac{1}{x \sigma \sqrt{2\pi}} \exp\left(-\frac{(\log(x) - \mu)^2}{2\sigma^2}\right), \quad x > 0
\]

Additionally, the probability that \( X \leq 1 \) is:

\[
P(X \leq 1) = P(log(X)\leq0)=P(Y\leq0)= P\left(Z \leq -\frac{\mu}{\sigma}\right)= \Phi\left(-\frac{\mu}{\sigma}\right)
\]
where \(\Phi\) denotes the cumulative distribution function of the standard normal distribution.

For our analysis, we fit a Log-Normal density function to the data obtained for different player sizes in Big Boss games. Specifically, for a game of size $n$, we perform a log-normal least squares fitting, yielding estimates for the parameters $\hat{\mu}_X$ and $\hat{\sigma}_X$. Furthermore, using these estimated parameters, we can compute the expected value and variance of our estimation using the following formulas:


\[
E[\hat{X}] = e^{\hat{\mu}_X + \frac{\hat{\sigma}_X^2}{2}}, \quad V[\hat{X}] = (e^{\hat{\sigma}_X^2} - 1)e^{2\hat{\mu}_X + \hat{\sigma}_X^2}
\]
where $\hat{X} \sim \text{Log-Normal}(\hat{\mu}_X, \hat{\sigma}_X^2)$. 

Table \ref{tab:big} shows the estimates for Big Boss games with up to 11 players, as we have been unable to solve the case with 12 players due to its computational complexity. Additionally, we set the seed \(\mu = 1000\) to increase the diversity of the selected Big Boss games in the samples and to demonstrate that this value does not alter predictions obtained with smaller values. From the obtained data, we can observe that the estimate of the mean increases with the size of the game, although the standard deviation seems to stabilize around a range of values.

%


\begin{table}[H]
\centering
\begin{tabular}{|c|c||c|c||c|c||c|c|}
$n$ & Sample size ($m$) & $\hat{\mu}_{X}$ & $\hat{\sigma}_{X}$ & $E[\hat{X}]$ & $V[\hat{X}]$ & $P(\hat{X} \leq 1)$ & $\frac{\#\{X_i \leq 1\}}{m}$\\ 
\hline
\hline
3 & 75000 & -0.63945 & 0.12035 & 0.53142 & 0.0041203 & 1 & 0.91085 \\ 
\hline
4 & 100000 & -0.36859 & 0.22352 & 0.70921 & 0.025768 & 0.95043 & 0.83398 \\ 
\hline
5 & 125000 & -0.20314 & 0.26217 & 0.8447 & 0.050766 & 0.78079 & 0.70969 \\ 
\hline
6 & 150000 & -0.080581 & 0.24528 & 0.95076 & 0.056054 & 0.62874 & 0.59314 \\ 
\hline
7 & 175000 & 0.019859 & 0.2414 & 1.0502 & 0.066182 & 0.46722 & 0.44794 \\ 
\hline
8 & 200000 & 0.095919 & 0.22983 & 1.1301 & 0.069274 & 0.33821 & 0.32211 \\
\hline
9 & 225000 & 0.16789 & 0.23043 & 1.2146 & 0.080455 & 0.23312 & 0.21786 \\ 
\hline
10 & 250000 & 0.20518 & 0.21246 & 1.2558 & 0.072813 & 0.16709 & 0.15492 \\
\hline
11 & 275000 & 0.24039 & 0.20261 & 1.2981 & 0.070612 & 0.11771 & 0.11036 \\
\hline
\end{tabular}
\caption{Log-Normal random variable parameter estimates}
\label{tab:big}
\end{table}

It is important to highlight that the probability of the distribution taking values less than or equal to one refers to the percentage of times that the PSV allocation, \( \tau^{\alpha^{v}}(v) \), lies between \( \tau(v) \) and \( e_1(v) \), or symmetrically, \( P(\hat{X} > 1) \) is equivalent to the percentage of times \( \tau^{\alpha^{v}}(v) = e_1(v) \). As we can observe in Table \ref{tab:big}, when the number of players increases, it becomes less likely that \( \tau^{\alpha^{v}}(v) \) differs from \( e_1(v) \).  

For this reason, we wonder at what number of players we can consider it almost certain that this occurs. To address this, we will study the evolution of the population parameter estimates for the Log-Normal random variable.

\begin{figure}[H]
    \centering
    \begin{minipage}{0.45\textwidth}
        \centering
        \begin{tikzpicture}
            \begin{axis}[
                xlabel={n},
                ylabel={$\hat{\mu}_{X}$},
                grid=both,
                xmin=2, xmax=11.5,
                ymin=-0.7, ymax=0.26,
                every axis plot/.append style={mark=*},  
            ]
	  \addplot coordinates {(3, -0.63945) (4,-0.36859 ) (5,-0.20314) (6,-0.080581) (7,0.019859) (8, 0.095919) (9,  0.16789) (10,0.20518 ) (11, 0.24039)};
            \end{axis}
        \end{tikzpicture}
        \caption{$\hat{\mu}_{X}$ estimates}
    \end{minipage}
    \hfill  
    \begin{minipage}{0.45\textwidth}
        \centering
        \begin{tikzpicture}
            \begin{axis}[
                xlabel={n},
                ylabel={$\hat{\sigma}_{X}$},
                grid=both,
                xmin=2, xmax=11.5,
                ymin=0.11, ymax=0.27,
                every axis plot/.append style={mark=*},  
            ]
	      \addplot coordinates {(3,0.12035) (4,0.22352) (5,0.26217) (6,0.24528) (7, 0.2414) (8,0.22983) (9,0.23043) (10,0.21246) (11, 0.20261)};
            \end{axis}
        \end{tikzpicture}
        \caption{\(\hat{\sigma}_{X}\) estimates}
    \end{minipage}
\end{figure}

As we can observe, the estimate of the parameter \(\hat{\mu}_{X}\) appears to follow a logarithmic trend, while \(\hat{\sigma}_{X}\) seems to stabilize within a specific range of values. When applying logarithmic regression to \(\hat{\mu}_{X}\), we obtain a coefficient of determination of \(R^2 = 0.9845\). For \(\hat{\sigma}_{X}\), we apply a moving average smoothing technique.

\begin{figure}[H]
    \centering
    \begin{minipage}{0.45\textwidth}
        \centering
        \includegraphics[width=\linewidth]{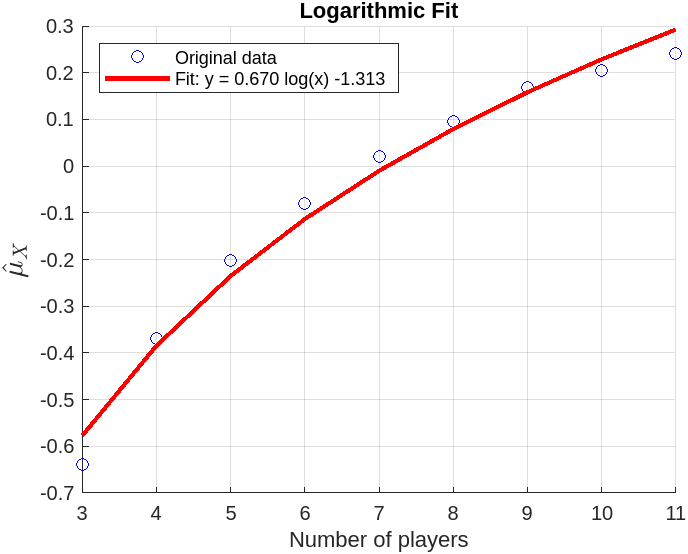}  
        \caption{Logarithmic regression for \(\hat{\mu}_{X}\) estimates}  
    \end{minipage}
    \hfill  
    \begin{minipage}{0.45\textwidth}
        \centering
        \includegraphics[width=\linewidth]{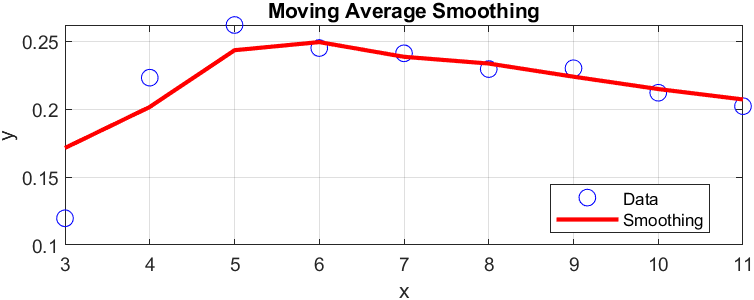}  
        \caption{Moving average smoothing of \(\hat{\sigma}_{X}\) estimates }  
    \end{minipage}

\end{figure}

According to the moving average smoothing for \(\hat{\sigma}_{X}\), we obtain a residual standard deviation of 0.021032 and a predicted value of 0.21517, which we use as \(\hat{\sigma}_{X}\), in our predictions. Additionally, to estimate $\hat{\mu}_{X}$, we use the regression equation:  \[\hat{\mu}_{X} = 0.67 \log(n) - 1.313.\] 
It is important to note that regression is generally not a method commonly used for making predictions outside the range of the data. However, given the high value of the coefficient of determination (\( R^2 = 0.9845 \)), we allow ourselves the flexibility to forecast the estimated parameters when slightly increasing the number of players.

Table \ref{tab:pre} presents the predictions for Big Boss games with 12 to 15 players. Given the strong logarithmic fit, we extend our analysis to explore what would happen in games with 15 players. As observed, the probability that \( \tau^{\alpha^{v}}(v) \) differs from \( e_1(v) \) is almost zero (\( P(\hat{X} \leq 1) = 0.00990 \)).

\begin{table}[h!]
\centering
\begin{tabular}{|c|c|c|c|}
$n$ & $\hat{\mu}_{X}$ & $\hat{\sigma}_{X}$ & $P(\hat{X} \leq 1)$  \\
\hline
 12 & 0.35189  &  0.21517  &0.05098 \\
\hline
 13 & 0.40552 &  0.21517  & 0.02974  \\
\hline
 14 &  0.45517  &  0.21517  &  0.01720    \\
\hline
 15 &  0.50139  &  0.21517  & 0.00990    \\
\hline
\end{tabular}

\caption{Log-Normal random variable parameter predictions  }
\label{tab:pre}
\end{table}

Thus, we can assert that the PSV allocation \( \tau^{\alpha^{v}}(v) \) will almost certain be equal to \( e_1(v) \) for games with a large number of players (\( n > 15 \)). In other words, the projection of the Shapley value onto the segment \( T(N,v) \) will be the endpoint \( e_1(v) \) without the need for explicit computation in large games.  

To conclude, this statistical analysis highlights the idea that strength lies in unity in Big Boss games. Specifically, as the number of weak players increases, their cumulative marginal contributions surpass those of the Big Boss, leading the Shapley value to compensate them significantly. Consequently, the projection of the Shapley value allows weak players to get closer to their maximum expectation $M_i(v)$ as their numbers grow. 

\section{Concluding Remarks} \label{con}
In this paper, we have revisited the class of games known as Big Boss games, delving into a fundamental part of the core of these games: the $\tau$-diagonal. This region of the core not only contains the allocations that are most relevant to both the Big Boss and the weak players, but also includes the $\tau$-value, which in this context acts as a solomonic allocation, representing the midpoint within the $\tau$-diagonal. In this framework, we have proposed a new allocation mechanism, the PSV allocation, which lies within the $\tau$-diagonal and captures the influence of the Shapley value by being its projection onto this set. We demonstrate that this new allocation benefits weak players, as it at least provides them with the same allocation as the $\tau$-value, while also considering their weight in the total benefit of the game. In particular, if their accumulated marginal contributions increase relative to the Big Boss, both the Shapley value and the PSV allocation provide them with a greater benefit.

Through the study of this allocation, we have identified a new property that describes the relationship between the increments of the $\tau$-value and the Shapley value, for both the Big Boss and the weak players. This property states that the increments experienced by the weak players can never exceed those of the Big Boss. From this relationship, we have proven that the PSV allocation is always closer to the maximum attainable claim (within the core) of the weak players. In fact, we have been able to characterize the convexity of the game in terms of the projection of the Shapley value onto the $\tau$-diagonal.

Since the calculation of the PSV allocation directly depends on the Shapley value, we conducted a statistical study to assess the feasibility of estimating the position of this allocation within the $\tau$-diagonal based on the size of the game. The results of the study show that such an estimation is possible, and that, when considering a sample of Big Boss games with a certain number of players, the distribution of the PSV allocation follows a log-normal distribution. This allows for relatively precise estimates of the position of the allocation within the $\tau$-diagonal. In particular, we conclude that, when the number of players is sufficiently large (more than 15), the probability that the PSV allocation grants the weak players their maximum aspiration (i.e., their marginal contribution to the grand coalition) approaches 1. This finding eliminates the need for complex computations in large games and highlights the impact of increasing the number of weak players.

Several future research directions arise from this work. One of them is to further explore the characterization of the PSV allocation, as well as extend these results to more general classes of games, such as Clan games. Furthermore, it would be interesting to conduct a more exhaustive statistical study that allows for greater precision in estimating the position of the PSV allocation and, consequently, provides an accurate estimation of the Shapley value. This line of research could be highly valuable in analyzing games with a large number of players. Finally, another potential research avenue would be to extend the results of this work to interval games, which would open up new opportunities for applying the developed techniques to more complex contexts.

\section*{Acknowledgements}
This work is part of the R+D+I project grants PID2022-137211NB-100,
that were funded by MCIN/AEI/10.13039/501100011033/ and by
‘‘ERDF A way of making Europe’’/EU. This research was also funded
by project PROMETEO/2021/063 from the Conselleria d'Innovació, Universitats, Ciència i Societat Digital, Generalitat Valenciana. The authors would like to thank Professor Domingo Morales for his insightful comments on this paper.

This paper is dedicated to the memory of Stef Tijs, for his contribution to the creation and consolidation of most of the cooperative game theory research groups in Spain. Thank you, Stef, for your dedication, for transmitting your vast knowledge, for your tireless commitment, and for the affection you always showed to the GATHER group at the I.U. Centro de Investigación Operativa of the Universidad Miguel Hernández de Elche. Your legacy lives on in each of us, and your influence on our research and our professional and personal lives endures.

\subsection*{Declarations}
\textbf{Conflict of interest} Not applicable.
 
\bibliography{bibliography}
\bibliographystyle{apalike}

 \newpage
\section*{Appendix - Proofs}

\begin{proof}[Proof of Theorem \protect\ref{dis}]
Take $v\in BBG^{N}$. If $M_{i}(v) = 0$ for all \( i \in N \setminus \{b^{v}\} \), the solution is trivial, as the $\tau$-diagonal reduces to a single point. For convenience, we will set $\alpha^{v} := 1$ in this case. Now, suppose that $\sum_{i \in N \setminus \{b^{v}\}} M_{i}(v) > 0$. To select the element of $T(N,v)$ that minimizes the distance to the Shapley value, it is sufficient to solve the following optimization problem:

\[
\displaystyle \min_{x\in T(N,v)}\left\{ \sqrt{\sum_{i\in N}\left[ x_{i}-\phi _{i}(N,v)\right]^{2}}\right\}
\]

The optimum of this problem will be the same as the one we
obtain in:
\[
\displaystyle \min_{x\in T(N,v)}\left\{ \sum_{i\in N}\left[ x_{i}-\phi _{i}(N,v)\right]^{2}\right\}
\] Finally, we can simplify the optimization
problem to obtain:
\begin{eqnarray*}
\min  &&f(\rho ):=\sum_{i\in N\setminus \{b^{v}\}}\left[ \rho M_{i}(v)-\phi
_{i}(v)\right] ^{2}+\left[ v(N)-\rho \sum_{i\in N\setminus
\{b^{v}\}}M_{i}(v)-\phi _{b^{v}}(v)\right] ^{2} \\
\mbox{s.t.} &&\rho \in \lbrack 0,1],
\end{eqnarray*}%
Since $f$ is a continuous and differentiable function, and the feasible set
is compact, we are guaranteed the existence of the optimum. Therefore, $%
f^{^{\prime }}(\rho )=0$ is equivalent to: 
\fontsize{10}{10}
\begin{gather}
2\sum_{i\in N\setminus \{b^{v}\}}\left[ \rho M_{i}(v)-\phi _{i}(v)\right]
\cdot M_{i}(v)-2\left[ v(N)-\rho \sum_{i\in N\setminus
\{b^{v}\}}M_{i}(v)-\phi _{b^{v}}(v)\right] \sum_{i\in N\setminus
\{b^{v}\}}M_{i}(v)=0; \notag\\
\rho \sum_{i\in N\setminus \{b^{v}\}}M_{i}^{2}(v)-\sum_{i\in N\setminus
\{b^{v}\}}\left( \phi _{i}(v)\cdot M_{i}(v)\right) +\rho \left( \sum_{i\in
N\setminus \{b^{v}\}}M_{i}(v)\right) ^{2}-\left( v(N)-\phi
_{b^{v}}(v)\right) \sum_{i\in N\setminus \{b^{v}\}}M_{i}(v)=0;  \notag \\
\rho \sum_{i\in N\setminus \{b^{v}\}}M_{i}^{2}(v)+\rho \left( \sum_{i\in
N\setminus \{b^{v}\}}M_{i}(v)\right) ^{2}=\left( v(N)-\phi
_{b^{v}}(v)\right) \sum_{i\in N\setminus \{b^{v}\}}M_{i}(v)+\sum_{i\in
N\setminus \{b^{v}\}}\left( \phi _{i}(v)\cdot M_{i}(v)\right) ;  \notag \\
\rho ^{v}=\frac{\left( v(N)-\phi _{b^{v}}(v)\right) \cdot \left(
\sum_{i\in N\setminus \{b^{v}\}}M_{i}(v)\right) +\sum_{i\in N\setminus
\{b^{v}\}}\left( \phi _{i}(v)\cdot M_{i}(v)\right) }{\left( \sum_{i\in
N\setminus \{b^{v}\}}M_{i}(v)\right) ^{2}+\sum_{i\in N\setminus
\{b^{v}\}}M_{i}^{2}(v)}\geq 0\label{foru}
\end{gather}%
\fontsize{10}{10}
We have that $f^{^{\prime \prime }}(\rho )=2\sum_{i\in N\setminus
\{b^{v}\}}M_{i}^{2}(v)+2\left( \sum_{i\in N\setminus
\{b^{v}\}}M_{i}(v)\right) ^{2}>0,$ then $f(\rho )$ it is convex, and since
the feasible set is bounded, the optimum $\alpha ^{v}$ is located at $\alpha
^{v}:=\min \{1,\rho ^{v }\}.$
\end{proof}
\medskip
\begin{proof}[Proof of Proposition \protect\ref{suf}]
Take $v\in BBG^{N}$. If $M_{i}(v) = 0$ for all \( i \in N \setminus \{b^{v}\} \), then we set $\alpha^{v} = 1$. Suppose now that $\sum_{i \in N \setminus \{b^{v}\}} M_{i}(v) > 0$. Consider the direction vector of the segment $T(N,v)$ is:
\begin{equation*}
\overrightarrow{v}:=\left\{ 
\begin{array}{ll}
M_{i}(v), & i\in N\setminus \{b^{v}\}, \\ 
&  \\ 
-\sum_{i\in N\setminus \{b^{v}\}}M_{i}(v), & i=b^{v},%
\end{array}%
\right. 
\end{equation*}%
Then,  the hyperplane that passes through \( \tau(v) \) and contains all the points in space whose projection onto \( T(N,v) \) is \( \tau (v) \) is given by 
\[
\pi : \left( -\sum_{i\in N\setminus \{b^{v}\}}M_{i}(v)\right) \left( x_{b^{v}}-\tau _{b^{v}}(v)\right) +\sum_{i\in N\setminus \{b^{v}\}}\left[ M_{i}(v)\left( x_{i}-\tau _{i}(v)\right) \right] =0.
\]  

Thus, the condition \( \alpha ^{v}\geq \frac{1}{2} \) is equivalent to:  

\[
\left( -\sum_{i\in N\setminus \{b^{v}\}}M_{i}(v)\right) \left(\phi _{b^{v}}(v)-\tau _{b^{v}}(v)\right) +\sum_{i\in N\setminus \{b^{v}\}}\left[ M_{i}(v)\left(\phi _{i}(v)-\tau _{i}(v)\right) \right] \geq 0;
\]  

\[
\sum_{i\in N\setminus \{b^{v}\}} M_{i}(v) \left(\phi _{i}(v)-\tau _{i}(v)\right) \geq \left( \sum_{i\in N\setminus \{b^{v}\}}M_{i}(v)\right) \left(\phi _{b^{v}}(v)-\tau _{b^{v}}(v)\right).
\]  

A sufficient condition for this to hold is that  

\[
\phi _{i}(v)-\tau _{i}(v) \geq \phi _{b^{v}}(v)-\tau _{b^{v}}(v), \quad \text{for all } i \in N\setminus \{b^{v}\}.
\]  

This is equivalent to:  

\[
\tau _{b^{v}}(v)-\phi _{b^{v}}(v) \geq \max_{i \in N\setminus \{b^{v}\}} \{ \tau _{i}(v)-\phi _{i}(v)\}.
\]  

\end{proof}
\medskip
\begin{proof}[Proof of Proposition \protect\ref{pa3}]
Take $v\in BBG^{N}$. Consider a coalition $S \subseteq N$ with $b^{v} \in S $. Then, by property (B3) we have that:
\begin{eqnarray*}
v(N)-v(S) &\geq &\sum_{i\in N\setminus S}M_{i}(v); \\
v(N)-v(S) &\geq &\sum_{i\in N\setminus S}\left[ v(N)-v(N\setminus \{i\})%
\right] ; \\
v(N)-v(S) &\geq &\left( n-s \right) v(N)-\sum_{i\in
N\setminus S}v(N\setminus \{i\}); \\
\left( n-s -1\right) v(N)+v(S) &\leq &\sum_{i\in
N\setminus S}v(N\setminus \{i\}).
\end{eqnarray*}
\end{proof}
\medskip
\begin{proof}[Proof of Theorem \protect\ref{te}]
Take $v\in BBG^{N}$. The inequality is trivially satisfied if $ M_{i}(v) = 0 = \tau_{i} (v)$ for all \( i \in N \setminus \{b^{v}\} \). Otherwise, take \( j \in N \setminus \{b^{v}\} \), we need to prove that \( \tau _{j}(v) - \phi _{j}(N,v) \leq \tau _{b^{v}}(v) - \phi _{b^{v}}(v) \). Each term of the inequality can be written as follows:

\scriptsize 
\begin{gather*}
\tau _{b^{v}}(v)-\phi _{b^{v}}(N,v)=v(N)-\frac{1}{2}\sum_{i\in N\setminus
\{b^{v}\}}\left( v(N)-v(N\backslash \{i\})\right) -\sum\limits_{S\subseteq
N\backslash \{b^{v}\}}\left[ \frac{s!(n-s-1)!}{n!}\cdot \left( v(S\cup
\{b^{v}\})-v(S)\right) \right]  \\
=v(N)-\frac{n-1}{2}\cdot v(N)+\frac{1}{2}\sum_{i\in N\setminus
\{b^{v}\}}v(N\backslash \{i\})-\frac{1}{n}v(N)-\sum\limits_{S\varsubsetneq
N\backslash \{b^{v}\}}\left[ \frac{s!(n-s-1)!}{n!}\cdot v(S\cup \{b^{v}\})%
\right]  \\
=\frac{3n-n^{2}-2}{2n}v(N)+\frac{n^{2}-n-2}{2n(n-1)}\sum_{i\in N\setminus
\{b^{v}\}}v(N\backslash \{i\})-\sum\limits_{\substack{ S\varsubsetneq
N\backslash \{b^{v}\} \\ \left\vert S\right\vert \leq n-3}}\left[ \frac{%
s!(n-s-1)!}{n!}\cdot v(S\cup \{b^{v}\})\right],
\end{gather*}%
\normalsize
and,
\scriptsize 
\begin{gather*}
\tau _{j}(v)-\phi _{j}(N,v)=\frac{1}{2}v(N)-\frac{1}{2}v(N\backslash
\{j\})-\sum\limits_{S\subseteq N\backslash \{j\}}\left[ \frac{s!(n-s-1)!}{n!}%
\cdot \left( v(S\cup \{j\})-v(S)\right) \right]  \\
=\frac{1}{2}v(N)-\frac{1}{2}v(N\backslash \{j\})-\frac{1}{n}v(N)+\frac{1}{n}%
v(N\backslash \{j\})-\sum\limits_{S\varsubsetneq N\backslash \{j\}}\left[ 
\frac{s!(n-s-1)!}{n!}\cdot \left( v(S\cup \{j\})-v(S)\right) \right]  \\
=\frac{n-2}{2n}v(N)-\frac{n-2}{2n}v(N\backslash
\{j\})-\sum\limits_{S\varsubsetneq N\backslash \{j\}}\left[ \frac{s!(n-s-1)!%
}{n!}\cdot \left( v(S\cup \{j\})-v(S)\right) \right]  \\
=\frac{n-2}{2n}v(N)-\frac{n-2}{2n}v(N\backslash
\{j\})-\sum\limits_{S\varsubsetneq N\backslash \{j,b^{v}\}}\left[ \frac{%
\left( s+1\right) !(n-s-2)!}{n!}\cdot \left( v(S\cup \{j,b^{v}\})-v(S\cup
\{b^{v}\}\right) \right]  \\
=\frac{n-2}{2n}v(N)-\frac{n-2}{2n}v(N\backslash \{j\})-\frac{1}{n(n-1)}\sum
_{\substack{ i\in N\setminus \{b^{v}\} \\ i\neq j}}v(N\backslash
\{i\})-\sum\limits_{\substack{ S\varsubsetneq N\backslash \{j,b^{v}\} \\ %
\left\vert S\right\vert \leq n-4}}\left[ \frac{\left( s+1\right) !(n-s-2)!}{%
n!}\cdot v(S\cup \{j,b^{v}\})\right] \\ +\sum\limits_{S\varsubsetneq
N\backslash \{j,b^{v}\}}\left[ \frac{\left( s+1\right) !(n-s-2)!}{n!}\cdot
v(S\cup \{b^{v}\})\right]. 
\end{gather*}%
\normalsize
Therefore, the inequality \( \tau _{j}(v) - \phi _{j}(N,v) \leq \tau _{b^{v}}(v) - \phi _{b^{v}}(v) \) is equivalent to:

\scriptsize 
\begin{gather*}
\\
\frac{n-2}{2}v(N)-\sum\limits_{\substack{ S\varsubsetneq N\backslash
\{j,b^{v}\} \\ \left\vert S\right\vert \leq n-4}}\left[ \frac{\left(
s+1\right) !(n-s-2)!}{n!}\cdot v(S\cup \{j,b^{v}\})\right]
+\sum\limits_{S\varsubsetneq N\backslash \{j,b^{v}\}}\left[ \frac{\left(
s+1\right) !(n-s-2)!}{n!}\cdot v(S\cup \{b^{v}\})\right]  \\
+\sum\limits_{\substack{ S\varsubsetneq N\backslash \{b^{v}\} \\ \left\vert
S\right\vert \leq n-3}}\left[ \frac{s!(n-s-1)!}{n!}\cdot v(S\cup \{b^{v}\})%
\right] \leq \frac{1}{2}\sum_{\substack{ i\in N\setminus \{b^{v}\} \\ i\neq j
}}v(N\backslash \{i\})+\frac{n-2}{n-1}v(N\backslash \{j\}); \\
\frac{n!\left( n-2\right) }{2}v(N)+\frac{n!}{n-1}v(\{b^{v}\})-\sum\limits
_{\substack{ S\varsubsetneq N\backslash \{j,b^{v}\} \\ \left\vert
S\right\vert \leq n-4}}\left[ \left( s+1\right) !(n-s-2)!\cdot v(S\cup
\{j,b^{v}\})\right] +\sum\limits_{\emptyset \neq S\varsubsetneq N\backslash
\{j,b^{v}\}}\left[ \left( s+1\right) !(n-s-2)!\cdot v(S\cup \{b^{v}\}\right]
 \\
\sum\limits_{\substack{ \emptyset \neq S\varsubsetneq N\backslash \{b^{v}\}
\\ \left\vert S\right\vert \leq n-3}}\left[ s!(n-s-1)!\cdot v(S\cup
\{b^{v}\})\right] \leq \frac{n!}{2}\sum_{\substack{ i\in N\setminus \{b^{v}\}
\\ i\neq j}}v(N\backslash \{i\})+\frac{n!\left( n-2\right) }{n-1}%
v(N\backslash \{j\}).
\end{gather*}%
\normalsize
Using the following equality:
\scriptsize
\begin{equation*}
\sum\limits_{\substack{ \emptyset \neq S\varsubsetneq N\backslash \{b^{v}\}
\\ \left\vert S\right\vert \leq n-3}}\left[ s!(n-s-1)!\cdot v(S\cup
\{b^{v}\})\right] =\sum\limits_{\substack{ \emptyset \neq S\subseteq
N\backslash \{j,b^{v}\} \\ \left\vert S\right\vert \leq n-3}}\left[
s!(n-s-1)!\cdot v(S\cup \{b^{v}\})\right] +\sum\limits_{\substack{ %
S\subseteq N\backslash \{j,b^{v}\} \\ \left\vert S\right\vert \leq n-4}}%
\left[ \left( s+1\right) !(n-s-2)!\cdot v(S\cup \{j,b^{v}\})\right] 
\end{equation*}%
\normalsize
We obtain:
\scriptsize 
\begin{gather*}
\frac{n!\left( n-2\right) }{2}v(N)+\frac{n!}{n-1}v(\{b^{v}\})+\sum\limits_{%
\emptyset \neq S\varsubsetneq N\backslash \{j,b^{v}\}}\left[ \left(
s+1\right) !(n-s-2)!\cdot v(S\cup \{b^{v}\})\right] + \\
\sum\limits_{\substack{ \emptyset \neq S\subseteq N\backslash \{j,b^{v}\} \\ %
\left\vert S\right\vert \leq n-3}}\left[ s!(n-s-1)!\cdot v(S\cup \{b^{v}\})%
\right] \leq \frac{n!}{2}\sum_{\substack{ i\in N\setminus \{b^{v}\} \\ i\neq
j}}v(N\backslash \{i\})+\frac{n!\left( n-2\right) }{n-1}v(N\backslash \{j\})
\end{gather*}%
\normalsize
Finally, using:
\scriptsize 
\begin{equation*}
\sum\limits_{\emptyset \neq S\varsubsetneq N\backslash \{j,b^{v}\}}\left[
\left( s+1\right) !(n-s-2)!\cdot v(S\cup \{b^{v}\})\right] +\sum\limits
_{\substack{ \emptyset \neq S\subseteq N\backslash \{j,b^{v}\} \\ \left\vert
S\right\vert \leq n-3}}\left[ s!(n-s-1)!\cdot v(S\cup \{b^{v}\})\right]
=\sum\limits_{\substack{ \emptyset \neq S\subseteq N\backslash \{j,b^{v}\}
\\ \left\vert S\right\vert \leq n-3}}\left[ ns!(n-s-2)!\cdot v(S\cup
\{b^{v}\})\right] 
\end{equation*}
\normalsize
The initial inequality, \( \tau _{j}(v) - \phi _{j}(N,v) \leq \tau _{b^{v}}(v) - \phi _{b^{v}}(v) \), can be written as:
\scriptsize 
\begin{gather}\label{ex1}
\frac{n!\left( n-2\right) }{2}v(N)+\frac{n!}{n-1}v(\{b^{v}\})+\sum\limits
_{\substack{ \emptyset \neq S\subseteq N\backslash \{j,b^{v}\} \\ \left\vert
S\right\vert \leq n-3}}\left[ n\cdot s!(n-s-2)!\cdot v(S\cup \{b^{v}\})%
\right] \leq \frac{n!}{2}\sum_{\substack{ i\in N\setminus \{b^{v}\} \\ i\neq
j}}v(N\backslash \{i\})+\frac{n!\left( n-2\right) }{n-1}v(N\backslash \{j\})
\end{gather}%
\normalsize

The next step is to group each $v(S)$ from the first part of the inequality with $v(N)$ according to the Big Boss games property, rewritten in Proposition \ref{pa3}. That is, for each $v(\{b^{v}\})$, we need $n-2$ instances of $v(N)$, while for each $v(S \cup \{b^{v}\})$, we need $n-s-2$ due to the number of possible coalitions of that size, which is $\binom{n-2}{s}$. Specifically:

\begin{eqnarray*}
&&\frac{n!}{n-1}(n-2)+n \sum\limits_{s=1}^{n-3}\left[ s!(n-s-2)!(n-s-2)%
\binom{n-2}{s} \right]  \\
&=& \frac{n!}{n-1}(n-2)+n \sum\limits_{s=1}^{n-3}\left[ s!(n-s-2)!(n-s-2)%
\frac{(n-2)!}{s!(n-s-2)!}\right]  \\
&=& \frac{n!}{n-1}(n-2)+n(n-2)! \sum\limits_{s=1}^{n-3}(n-s-2) \\
&=& \frac{n!}{n-1}(n-2)+n(n-2)! \left[ (n-2)(n-3)-\frac{(n-3)(n-2)}{2}\right]  \\
&=& \frac{n!}{n-1}(n-2)+n(n-2)! \frac{(n-3)(n-2)}{2} =\frac{n!}{n-1}(n-2)+n! \frac{(n-3)(n-2)}{2(n-1)}\\
&=& \frac{n!(n-2)}{2}\left[ \frac{2}{n-1}+\frac{(n-3)}{n-1}\right] =\frac{n!(n-2)}{2}.
\end{eqnarray*}

This demonstrates that we can combine the coefficient of \( v(N) \) from the first part of the inequality with the coefficients of \( v(S) \) in a manner that ensures all combinations result in expressions of the form \( \left( n - s - 1 \right) v(N) + v(S) \), each multiplied by a specific factor. Formally, the first part of inequality (\ref{ex1}) can be rewritten as follows:

\small
\begin{equation*}
\frac{n!}{n-1}\left[ (n-2)v(N)+v(\{b^{v}\})\right] +\sum\limits_{\substack{ %
\emptyset \neq S\subseteq N\backslash \{j,b^{v}\} \\ \left\vert S\right\vert
\leq n-3}}\left[ n\cdot s!(n-s-2)!\cdot \left[ (n-s-2)v(N)+v(S\cup \{b^{v}\})%
\right] \right] 
\end{equation*}
\normalsize

Thanks to this, we can determine the number of such expressions and apply the result from Proposition \ref{pa3} to each of them. The reader may observe that both $\{b^{v}\}$ and $S \cup \{b^{v}\}$ do not include player $j$, which implies that every expression formed will contribute a summand $v(N \backslash \{j\})$ when the inequality is applied. The total number of such expressions is $\frac{n!}{n-1}$ for $S = \{b^{v}\}$ and  
$n \sum\limits_{s=1}^{n-3} \left[ s!(n-s-2)!\binom{n-2}{s} \right]$  
for $S \subseteq N \backslash \{j, b^{v}\}$ with $s \leq n-3$. Therefore, we obtain:

\begin{eqnarray*}
&&\frac{n!}{n-1}+n \sum\limits_{s=1}^{n-3}\left[ s!(n-s-2)! \binom{n-2}{s} \right] = \frac{n!}{n-1} + n (n-2)! \sum\limits_{s=1}^{n-3} 1  \\
&=& \frac{n!}{n-1} + n (n-2)! (n-3)=\frac{n!}{n-1} + n (n-2)! (n-3) = \frac{n!}{n-1} + \frac{n!(n-3)}{n-1}  \\
&=& \frac{n!}{n-1}(1 + (n-3)) = \frac{n!(n-2)}{n-1}
\end{eqnarray*}
The term \( v(N \backslash \{j\}) \) appears exactly as many times as dictated by expression (\ref{ex1}). Finally, we need to determine the number of occurrences of $v(N \backslash \{i\})$ with $i \neq j$ we can obtain. As in the previous case, we obtain $\frac{n!}{n-1}$ for $S=\{b^{v}\}$ and we must take into account that coalitions of the form $S \subseteq N \backslash \{j, b^{v}\}$ containing player $i$ will not produce any $v(N \backslash \{i\})$. Therefore, in this case, we will have $n \sum\limits_{s=1}^{n-3}\left[ s!(n-s-2)!\binom{n-3}{s} \right]$ for $S \subseteq N \backslash \{j, b^{v}\}$ with $s \leq n-3$. Summing both quantities, we get:
\small
\begin{eqnarray*}
&&\frac{n!}{n-1} + n \sum\limits_{s=1}^{n-3} \left[ s!(n-s-2)! \binom{n-3}{s} \right] = \frac{n!}{n-1} + n \sum\limits_{s=1}^{n-3} \left[ s!(n-s-2)! \frac{(n-3)!}{s!(n-s-3)!} \right]  \\
&=& \frac{n!}{n-1} + n (n-3)! \sum\limits_{s=1}^{n-3} (n-s-2) = \frac{n!}{n-1} + n (n-3)! \left[ (n-2)(n-3) - \frac{(n-3)(n-2)}{2} \right]  \\
&=& \frac{n!}{n-1} + n (n-3)! \frac{(n-3)(n-2)}{2} = \frac{n!}{n-1} + \frac{n!(n-3)}{2(n-1)} = \frac{2n! + n!(n-3)}{2(n-1)} = \frac{n!(n-1)}{2(n-1)} = \frac{n!}{2}
\end{eqnarray*}
\normalsize

It can be verified that the number of occurrences also matches the count given in expression (\ref{ex1}) after applying the inequality from Proposition \ref{pa3}, thereby proving the required inequality.

\end{proof}
\medskip
\begin{proof}[Proof of Corollary \protect\ref{col}]
Immediate upon applying Theorem \ref{te} and Proposition \ref{suf}.

\end{proof}
\medskip
\begin{proof}[Proof of Proposition \protect\ref{conv}]
($\Longrightarrow$) If $v$ is convex, it follows from Theorem \ref{muto} that $\phi(v) = \tau(v) = \tau^{1/2}(v)$, and hence, $\alpha^v = \frac{1}{2}$.

($\Longleftarrow$) If $\alpha^v = \frac{1}{2}$, then $\phi(v)$ lies in the hyperplane that passes through $\tau(v)$ and contains all the points in the space whose projection onto $T(N, v)$ is $\tau(v)$. By a construction similar to the beginning of the proof of Proposition \ref{suf}, we have that:

\[
\left( \sum_{i \in N \setminus \{b^{v}\}} M_i(v) \right) \left( \tau_{b^{v}}(v) - \phi_{b^{v}}(v) \right) = \sum_{i \in N \setminus \{b^{v}\}} M_i(v) \left(  \tau_i(v) - \phi_i(v)\right).
\]

Furthermore, by Proposition \ref{pa3}, we know that $\tau_{b^{v}}(v) - \phi_{b^{v}}(v) \geq \max_{i \in N \setminus \{b^{v}\}} \{ \tau_i(v) - \phi_i(v) \}$.

If \( M_i(v) > 0 \) for all \( i \in N \setminus \{b^{v}\} \), it follows that \( \tau_{b^{v}}(v) - \phi_{b^{v}}(v) = \tau_i(v) - \phi_i(v) \) for all \( i \in N \setminus \{b^{v}\} \), and by the efficiency property, we conclude that \( \phi(v) = \tau(v) \). Finally, by Theorem \ref{muto}, the game is convex.



\end{proof}

\end{document}